\documentclass{aa}
\usepackage{amssymb,epsfig,graphicx}
\usepackage{txfonts}
\begin{document}
   \title{Spectroscopy and multiband photometry of the afterglow of intermediate duration $\gamma$-ray burst 040924 and its host galaxy  }

   \author{K. Wiersema \inst{1}, 
           A.~J.~van der Horst \inst{1}, 
           D.~A.~Kann \inst{2}, 
           E.~Rol \inst{3}, 
	   R.~L.~C.~Starling \inst{3,1}, 
	   P.~A.~Curran \inst{1},
	   J.~Gorosabel \inst{4}, 
	   A.~J.~Levan \inst{5}, 
	   J.~P.~U.~Fynbo\inst{6}, 
           A.~de~Ugarte~Postigo\inst{4}, R.~A.~M.~J.~Wijers \inst{1},
	   A.~J.~Castro-Tirado \inst{4}, 
	   S.~S. Guziy \inst{4}, 
	   A.~Hornstrup \inst{7},
	   J.~Hjorth \inst{6},
	   M.~Jel\'{i}nek \inst{4},
	   B.~L.~Jensen \inst{6},
	   M. Kidger \inst{8},
	   F.~Mart\'{i}n-Luis \inst{9},
	   N.~R.~Tanvir \inst{3}, 
	   P.~Tristram \inst{10},
	   P.~M.~Vreeswijk \inst{11}
	   }

   \offprints{K. Wiersema}

   \institute{Astronomical Institute ``Anton Pannekoek'', University of Amsterdam,
              Kruislaan 403,  1098 SJ Amsterdam, The Netherlands. \\
              \email{kwrsema@science.uva.nl}
         \and
             Th\"uringer Landessternwarte Tautenburg,  Sternwarte 5, D--07778 Tautenburg, Germany.
	  \and   
	     Department of Physics and Astronomy, University of Leicester, University Road, Leicester, LE1 7RH, United Kingdom.
         \and
             Instituto de Astrof\'{i}sica de Andaluc\'{i}a (IAA-CSIC), Apartado de Correos, 3004, 18080 Granada, Spain.
         \and 
             Department of Physics, University of Warwick, Coventry, CV4 7AL, United Kingdom.
	  \and
	     Dark Cosmology Centre, Niels Bohr Institute, University of Copenhagen, Juliane Maries Vej 30, DK-2100 Copenhagen, Denmark.
	  \and
	     Danish National Space Center, Juliane Maries Vej 30, Copenhagen 0, DK-2100, Denmark.   
	  \and
	     ESA-European Space Astronomy Centre, Villafranca del Castillo, P.O. Box 50727, 28080 Madrid, Spain
	  \and
	     Instituto de Astrof\'{i}sica de Canarias, C/. Via Lactea s/n, 38205 La Laguna, Tenerife, Spain 
	  \and   
	      University of Canterbury, Private Bag 4800 Christchurch 8140, New Zealand.
	  \and
	     European Southern Observatory, Alonso de C\'{o}rdova 3107, Casilla 19001, Santiago 19, Chile.
               }

\date{Received 2007; accepted 2007}

\abstract
{}
   {We present optical photometry and spectroscopy of the afterglow and host galaxy of gamma-ray burst 040924. 
   This GRB had a rather short duration of $T_{90} \sim2.4$s, and a well sampled optical afterglow light curve. We aim to use this
   dataset to find further evidence that this burst is consistent with a massive star core-collapse progenitor. }
   {We combine the afterglow data reported here with data taken from the literature
   and compare the host properties with survey data. }
   {We find that the global behaviour of the optical afterglow is well fit by a broken power-law, with a break at $\sim$0.03 days.  
   We determine the redshift $z = 0.858 \pm 0.001$ from the detected emission lines in our spectrum. Using the spectrum and photometry 
   we derive global properties of the host, showing it to have similar properties to long GRB hosts.
   We detect the \ion{[Ne}{III]} emission line in the spectrum, and compare the fluxes of this line of a sample of 15 long GRB host galaxies with
   survey data, showing the long GRB hosts to be comparable to local metal-poor emission line galaxies in their \ion{[Ne}{III]} emission.
   We fit the supernova bump accompanying this burst, and find that it is similar to other long GRB supernova bumps, but fainter. }
   {All properties of GRB\,040924 are consistent with an origin in the core-collapse of a massive star: the supernova, the spectrum and SED of 
   the host and the afterglow. }


   \titlerunning{Spectroscopy and multiband photometry of gamma-ray burst 040924}
   \authorrunning{Wiersema et al.}
   \maketitle
%
\section{Introduction}
The large sample of $\gamma$-ray bursts (GRBs) detected by the BATSE instrument on board the CGRO satellite clearly shows that GRBs can be 
classified in two distinct classes, based on spectral hardness and duration (Kouveliotou et al.~1993), though the possibility of three 
classes has also been proposed (see e.g. Horv{\'a}th et al.~2006; see also Hakkila et al.~2003).
The difference in duration and spectral hardness 
may reflect important differences in the production mechanism of relativistic jets by the progenitor object, and therefore may be an important clue to the
nature of the progenitor. 
The great difficulty in obtaining accurate $\gamma$-ray positions for the short-hard burst population made searches for afterglows unsuccessful for a long time. 
With the arrival of  Swift and its ability to rapidly identify X-ray afterglows also for faint and short duration bursts, this situation changed 
dramatically, resulting in several localisations of short bursts by both Swift and HETE-2, and the detection of X-ray and optical 
afterglows for a considerable fraction of these. 
With the different sensitivity and spectral range that is covered by these satellites with respect to BATSE, the classification of
bursts in the classes of short and long burst became more ambiguous in several cases, for example due to long lasting soft tails detected in the prompt
emission of bursts that BATSE would have classified as members of the short-hard class (see e.g. Norris \& Bonnell~2006 and references therein). 
The question of the progenitors of short bursts versus long bursts can now also be approached through other angles, such as the presence of a supernova, the 
nature of the host galaxies of a
large sample of bursts, the density in which the afterglow propagates and its gradient, the circumburst environment through afterglow spectroscopy to name just a few 
(see Lee \& Ramirez-Ruiz 2007; Nakar 2007 and references therein). The distinction between
short and long GRBs is under strong debate. A classification just by the duration
of the prompt emission has proved to be weak to identify progenitor models (see also
the recent
detection of two long bursts without supernovae, Fynbo et al.~2006b; Gal-Yam et al.~2006; Della Valle et al.~2006;
Gehrels et al.~2006). A large sample of
datasets
covering prompt emission, afterglow and host galaxies of GRBs at both sides
of the classical 2 second limit is required.
In this paper we focus on the properties of GRB\,040924. This burst has a duration very close to the classical 2 second long-short divide, 
and has been considered a short burst candidate (e.g. Fan et al.~2005).  

GRB\,040924 was localised by the Fregate and WXM instruments 
onboard the HETE-2 satellite on 2004 September 24 at 11:52:11 UT (Fenimore et al.~2004), and also detected by satellites within the
IPN network (Golenetskii et al.~2004). 
The burst was classified as an X-ray rich burst, based on the ratio of the fluences of the burst in 
the 7-30 keV and the 30-400 keV bands.
Donaghy et al.~(2006) report a T$_{90}$ duration of 2.39 $\pm$ 0.24 seconds from the HETE-2 data, and show that the duration alone would make this 
burst a good candidate for membership of the short GRB class. However, a clear spectral lag is present and the spectrum is soft, favouring a 
long burst class membership (Donaghy et al.~2006). As such it is an interesting burst to study.

In this paper we present our dataset on the afterglow and host GRB\,040924, supplemented with observations from literature. 
In \S2 we describe our observations. In \S3 we analyze the afterglow data, and in \S4 the data on the host galaxy. In \S5 we derive general properties for the
associated supernova in relation to other GRB related supernovae, and in \S6 we present our conclusions and discussion.

\section{Observations and analysis}

\subsection{Afterglow optical photometry}
Due to the quick localisation by HETE-2, the Palomar 60 inch telescope was able to rapidly find a candidate for the optical afterglow associated with 
this burst (Fox \& Moon 2004, Soderberg et al.~2006).
We observed the afterglow of GRB~040924 at several different observatories. All 
data were reduced with standard {\sc IRAF} and {\sc MIDAS} packages. Cosmic ray hits were removed using
the L.A.Cosmic software by Van Dokkum (2001). Astrometric calibration was done using the USNO-B1.0 catalogue.
Photometry was done using the {\sc IRAF} DAOPHOT and APPHOT
packages. The V, R and I band images were calibrated relative to the photometric sequence provided
by Henden (2004). 
The log of photometric observations and their results is shown in  
Table \ref{obstable}. 

We imaged the error box with the 0.6 m telescope at Mt. John Observatory (+ MOA camera) in the wide R-band filter.
A 300-s single image was obtained through clouds at 0.087 days after burst, but did not detect the afterglow.
Further imaging at 0.2 days did marginally detect the afterglow.
Two epochs of V, R and I band imaging of the afterglow position 
were done with the
Very Large Telescope (VLT) in Chile, using the Focal Reducer - Low Dispersion Spectrograph (FORS1).
The conditions during the first epoch were excellent (seeing $\sim$0.6 arcsec), but seeing increased in the second epoch.
Two afterglow observations were done with the 2.56-m Nordic Optical Telescope (NOT) at La Palma. 
Observations were carried out using the Andaluc\'{i}a Faint Object Spectrograph and Camera (ALFOSC).  
We performed an H-band observation 
using the near-infrared MPI f\"{u}r Astronomie General-Purpose Infrared Camera (MAGIC) on the 
1.5m telescope at Observatorio de Sierra Nevada, which we calibrated using 2MASS field stars.

\begin{figure}[h]
\centering
\includegraphics[width=9cm]{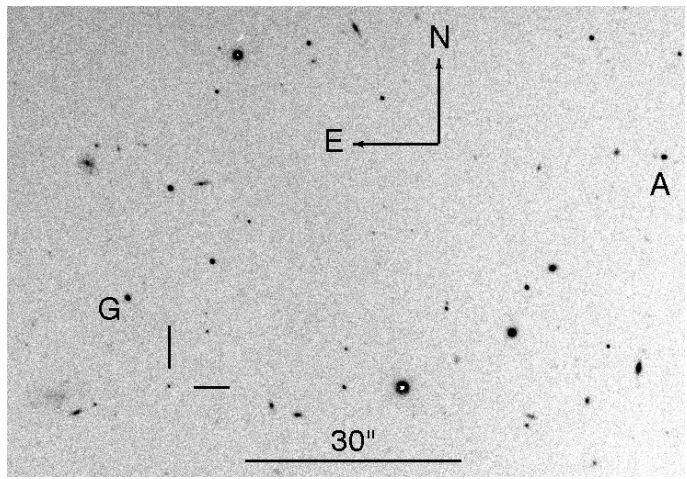}
\caption{Cut from the first epoch VLT FORS2 I band observation, showing the afterglow marked by tickmarks.
The star marked by {\em G} was used to align the FORS2 slit for the host galaxy spectroscopy. The moving object
marked by {\em A} is a main belt asteroid which was present in the first epoch V, R and I band observations.}
         \label{fcspec}
\end{figure}

\subsection{Spectroscopy}
Attempts at measuring the redshift from afterglow spectroscopy using the William Herschel Telescope (WHT) on La Palma on September 25 and 26
unfortunately proved unsuccessful due to bad weather decreasing the signal to noise ratio.
Therefore we obtained
a dark time spectrum using VLT UT1 (Antu) with the FORS2 
instrument in long slit spectroscopy (LSS) mode on October 10, 
approximately 15.75 days after burst. 
A bright star marked {\em G} in Figure \ref{fcspec} was used to position the slit, 
27 arcseconds away from the host galaxy, 
the position angle was approximately 25.1 degrees. 
Two spectra of each 1800 seconds exposure time were taken, using the
300V and 300I grisms, to increase wavelength coverage. 
We used a 1 arcsec slit.
Both exposures were done without an order sorting filter to increase sensitivity, which means that the red end of the 300V is contaminated by the second order 
(that has twice the resolution and wavelength). As there are no lines visible at the bluest region and the continuum is undetected,
this has no influence on the measured emission line fluxes.
Both spectra have been reduced in a standard fashion using the data reduction package 
{\sc IRAF}.
The L.A.Cosmic program (specifically the {\em lacos\_spec} routine) written by
Van Dokkum (2001) was used to remove both point- and irregular shaped
cosmic ray hits from the science and standard images.
The wavelength resolution is estimated by measuring the FWHM of the arc lines
at 10.2 \AA~for the 300V, and 9.6 \AA~for the 300I spectrum.
The spectra were flux calibrated by using the standard star LTT1788. 
Atmospheric extinction correction was done by applying a mean extinction
curve for Paranal. A Galactic extinction correction was performed by using the 
 $E(B-V)$ value of 0.058 mag (Schlegel, Finkbeiner \& Davis \cite{Schlegel98}), assuming a Galactic extinction law
$A_{\lambda}/A_{V}$ expressed as $R_{V} = A_{V}/E(B-V)$ (Cardelli et al.~1989). We make the standard assumption $R_{V} = 3.1$.

\begin{figure}[h]
\centering
\vspace{0.5cm}
\includegraphics[width=7.8cm]{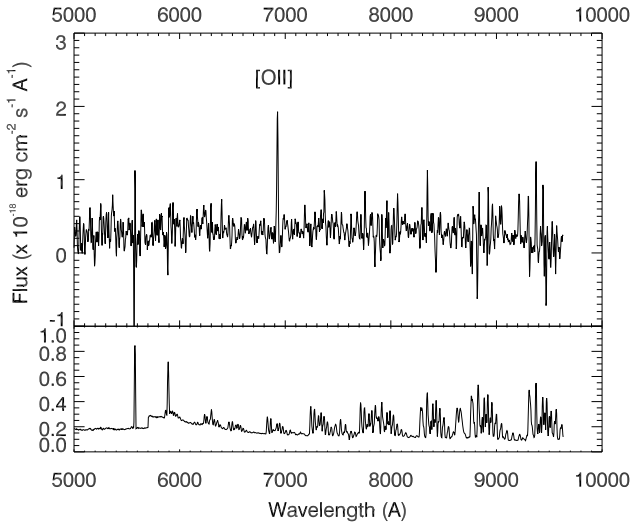}\vspace{0.8cm} 
\caption{The averaged VLT FORS2 300V and 300I spectrum of the host galaxy (upper panel), with corresponding error spectrum (lower panel). 
The prominent \ion{[O}{II]} line is labelled. The spectrum is smoothed by 3 pixels for displaying purposes.}
         \label{spectrum300IV}
\end{figure}

The combined spectra and the associated error spectrum are shown in Figure \ref{spectrum300IV}.
The continuum level of the host galaxy is faint, with a mean signal to noise ratio of $\sim$2 per pixel. 
We find several emission lines in the spectra, which we present in table \ref{linefluxes}.
All of these lines are consistent with a redshift $z$ = 0.858 $\pm$ 0.001. 
No other faint, galaxy-like sources, visible in the WHT imaging data near the GRB host galaxy, fell in the slit.

\subsection{Radio observations\label{secradio}}
The radio observations reported in this paper are all done using the Westerbork Synthesis Radio Telescope (WSRT) in 
the Netherlands. Three observations at 
the position of the afterglow were made, all at 4.9 GHz. 
Data reduction was performed using the MIRIAD software (Sault, Teuben \& Wright 1995).
No significant detection of an afterglow was found in any of the three epochs, see Table \ref{obstable}. 
To search for a radio host galaxy, we combine the data of all three observations. Again no radio source is found on 
the position of the afterglow, with corresponding 3$\sigma$ upper limit of 63 $\mu$Jy. A formal 
flux measurement for a point source at the location of the optical afterglow gives a value
of 12 $\pm$ 21 $\mu$Jy for the combined data.    

\subsection{Supernova search and host galaxy \label{snsearch}}
The observations reported in this section were aimed at observing a possible supernova associated 
with GRB\,040924, which was later identified through much deeper HST imaging (Soderberg et al.~2006). 
A redshifted, K-corrected extrapolation of supernova 1998bw would yield expected SN magnitudes of
R $\sim$24.6 and I $\sim$23.8, see section \ref{SNsection}, making it possible for us to detect it with 
ground based telescopes. 
Observations were done with the Auxiliary Port Imaging Camera (Aux-Port Imager) mounted 
at the auxiliary port of the Cassegrain A\&G box of the William Herschel Telescope (WHT).
The first epoch was done on October 23, consisting of 
4 x 15 minutes in I and 4 x 15 minutes in R. 
Whilst at the start of observations the conditions were good, the seeing greatly deteriorated during the 
exposures, which, combined with an increasing airmass, led to a relatively shallow limiting magnitude and low image quality of 
the combined R band exposure. 

A second epoch of I band imaging was performed on January 10, aimed at observing the host galaxy. 
The observations were done in the same manner as the first epoch, consisting of 4 $\times$ 15 minutes exposure time. The host galaxy is clearly detected. 
A second epoch of R band imaging was performed on July 16 2005, consisting of 3 $\times$ 15 minutes.

In order to get a broadband colour estimate of a potential supernova, two  K$^\prime$  band observations were done near the supernova peak
with the Omega2000 near-infrared wide field camera on the prime focus of the 3.5m telescope at the Calar Alto observatory, on 2004
October 26 and 27, with exposure times 2760 and 6390 seconds respectively. The host and SN are not detected. 

We note here that the photometric calibration of the WHT data are hampered by the very small field of view (the Aux Port imager
has a circular field of view of 1.8 arcmin diameter, but dithering required for fringe correction reduced the useful field of view to approx 1.2 arcmin). 
No Henden (2004) calibration stars can be used to calibrate, as there are only two of these in the field of view, and they are saturated. 
We use the NOT data to calibrate field objects,
but most objects near the GRB host that are in both NOT and WHT data are galaxies. The I band WHT data provided a better calibration this way than the R data, so
we point out that particularly the R band magnitudes of the WHT data should be treated with caution.

\begin{table*}[h]
\begin{center}
\begin{tabular}{llll}
\hline
 $\Delta$t (d) & Magnitude, Error (1$\sigma$) & Band / frequency & Telescope\\
\hline
0.204 & 21.3 $\pm$ 0.5 & R & MOA\\
0.568 & Upper limit 18.2 (3$\sigma$) & H & OSN MAGIC\\
0.666 &  22.97 $\pm$ 0.09 & V & VLT FORS2  \\
0.674 & 22.56 $\pm$ 0.13 & R & VLT FORS2 \\
0.679 & 21.95 $\pm$ 0.19 & I & VLT FORS2 \\
0.731 & 22.59 $\pm$ 0.15 & R & NOT ALFOSC\\
0.871 & 23.27 $\pm$ 0.08 & V & VLT FORS2 \\
0.878 & 22.85 $\pm$ 0.13 & R & VLT FORS2 \\
0.890 & 22.08 $\pm$ 0.19& I & VLT FORS2  \\
2.97 & 23.38 $\pm$ 0.29  & R & NOT ALFOSC\\
28.69 & 24.1 $\pm$ 0.30 & R & WHT Aux   \\
28.64 & 23.3  $\pm$  0.2& I & WHT Aux \\
31.72 & Upper limit 19.8 (3$\sigma$) & K$^\prime$ & CAHA Omega2000 \\
32.58 & Upper limit 19.6 (3$\sigma$) & K$^\prime$ & CAHA Omega2000 \\
108.39& 23.6 $\pm$ 0.2  & I & WHT Aux  \\
295.69 & 24.39 $\pm$ 0.30 &  R & WHT Aux  \\ 
\hline
0.663  & Flux $<$ 96 $\mu$Jy (3$\sigma$) & 4.9 GHz (6cm) & WSRT \\
5.541 & Flux $<$ 81 $\mu$Jy (3$\sigma$)&4.9 GHz (6cm)& WSRT \\
32.48 & Flux $<$ 75 $\mu$Jy (3$\sigma$) & 4.9 GHz (6cm) & WSRT \\
      & Flux $<$ 63 $\mu$Jy (3$\sigma$) &  4.9 GHz (6cm) & WSRT combined data\\ 
\hline
\end{tabular}
\caption{Log of observations of GRB\,040924 reported in this paper.\label{obstable} We note that the last NOT exposure is close to the limiting magnitude.}
\end{center}
\end{table*}

\section{The afterglow}
\label{afterglow}
\begin{figure}
  \centering
  \includegraphics[width=8.8cm, angle=0]{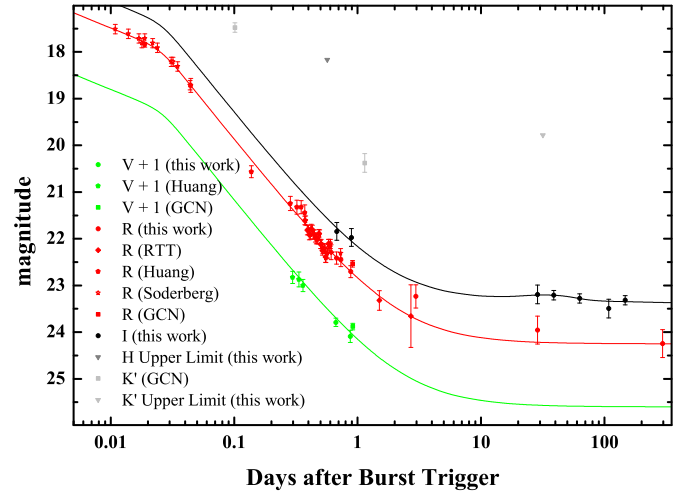}
  \caption{Light curves of the afterglow of GRB 040924, fitted with an achromatic broken power-law. The top curve is the 
  I band lightcurve and the middle curve is R band. The lower curve shows the V band curve where 1 magnitude is added for clarity.
  Data are from this paper and 
  sources given in \S \ref{afterglow}. The I band fit includes the contribution of a supernova.}
         \label{lightcurve}
   \end{figure}

We gathered published data from the literature 
(Terada et al.~2004a,b;  Silvey et al. 2004; Huang et al.~2005; Soderberg et al.~2006) 
and combined these with our data, also adding data from the RTT 150 telescope (I. Khamitov, priv. comm.\footnote{{\tt http://hea.iki.rssi.ru/$\sim$rodion/040924}}; 
Khamitov et al.~2004). We corrected the afterglow magnitudes for Galactic extinction where needed. The resulting light curve is shown in Fig.~\ref{lightcurve}.

We used the fitting method described in Curran et al.~(2007) which fits the data from all available bands 
simultaneously, assuming that each band has the same power-law decay but leaving free the offset between bands to account for differences
in filter curves and calibration. We refer to Curran et al.~(2007)
for details on the fitting method and error analysis. 
Our fit using all available data shows a broken power-law, with values of the afterglow decay index of $\alpha_1 \sim 0.31$, $\alpha_2 = 1.25 \pm 0.02$ and 
a break time of $\sim0.023$ days, where the break smoothness is fixed at a value $s=10$, with a best fit $\chi^2 / {\mbox{dof}}= 1.2$. We note that the
pre-break decay ($\alpha_1$) and the break time are not well constrained, the $\chi^2$ is fairly insensitive to changes in these parameters.  
The $R$ band afterglow shows a low significance wiggle around the best fit power-law, most noticeably around $\sim 0.5$ days. 
This is commonly seen in well sampled afterglow lightcurves, and does not affect our afterglow fit noticeably.

Using the data set presented in this paper, Kann et al. (2006) found that the VRIK$^\prime$ spectral energy distribution (SED) 
of the afterglow of GRB 040924 is fit well by a power-law plus SMC-like extinction, though the data are also well fit with no extinction. 
The extinction-corrected spectral slope is $\beta=0.63\pm0.48$ and the host 
galaxy extinction is $A_V=0.16\pm0.44$, agreeing with the value of the spectral slope $\beta = 0.61 \pm 0.08$ reported 
by Silvey et al.~(2004), who do not fit for extinction. 
A comparison with 
other intrinsic GRB afterglow luminosities (Kann et al. 2006) shows that the afterglow of GRB 040924 was the faintest in their sample of
pre-Swift GRB afterglows with observed optical counterparts in multiple bands and enough sampling to derive temporal and spectral indices.
Soderberg et al.~(2006) have previously noted the break from a shallow decay to a more common temporal decay index, and our decay slopes and break time 
agree with theirs, but improve on accuracy particularly for $\alpha_2$.  
This break is not consistent with a jet break, as the difference in temporal decay index is not as expected from a jet break, the break is very early, and both 
pre- and post-break decay slopes are too shallow (cf. Zeh et al.~2006 for a large sample of optical afterglow parameters). 
Huang et al.~(2005) reach similar results qualitatively, but have very different values due to a much smaller dataset.
The early data showing this break have only been taken in R band so it is unclear whether this break is achromatic, but after the early break 
multi-color data are available. The case of the synchrotron cooling and peak frequency redward of the observed optical frequency after the break, i.e. 
$\nu_{\rm m} < \nu_{\rm c} < \nu_{\rm opt}$, 
is ruled out, as the spectral and temporal slope can not be accomodated together in the standard blastwave model (e.g. Zhang \& M{\'e}sz{\'a}ros 2004).
The case for $\nu_{\rm m} < \nu_{\rm opt} < \nu_{\rm c}$ using a homogeneous density medium or a $1/{r^2}$ (stellar wind) medium would give $p \sim2.7 $ for 
a wind medium and $p \sim2.1$ for a homogeneous medium from the temporal decay. The spectral slope gives $p \sim2.2$ in this case. 
The distinction between the homogeneous and wind medium can be made by assuming $n(r) \propto r^{-k}$, and calculating $k$ using $\beta$ and $\alpha$
(we refer to Starling et al.~2007 for details). For the post-break values we find $k = 1.6^{+0.3}_{-0.5}$, which agrees with an interpretation of
a $1/{r^2}$ wind medium in which the blastwave propagates, though the uncertainties are quite large. 

The change in decay slope of $\Delta\alpha \sim0.9$ rules out the possibility that the break is caused by the passage of the cooling break $\nu_c$, for
which $\Delta\alpha = 0.25$ would be expected. Possibly this shallow decay phase and break can be attributed to 
energy injection, a mechanism frequently invoked to explain shallow decay phases in afterglows (e.g. Nousek et al.~2006; Pandey et al.~2006).
Using a parametrization $E \propto t^q$ (Nousek et al.~2006), we find $q\sim0.7$ in the case of a homogeneous medium, and $q\sim1.1$ in the 
case of a wind medium.

\section{The host galaxy \label{hostspec}}
We detect several emission lines in the VLT FORS2 spectra of the host (see figure \ref{spectrum300IV}), though the detections 
are of low significance due to the faintness of the host.
The analysis of the emission lines was done using the Starlink {\sc DIPSO} spectral fitting package, using mainly
the ELF (emission line fitting) routines, and compared with results using
the {\sc IRAF} splot package. The results were in agreement within errors. 
We note that several of the emission lines (particularly the \ion{[O}{III]}$\lambda$5007) are
affected by skylines and show differing fluxes in the 300V and 300I spectra, caused by the relative success of skyline
subtraction. We have not included these lines in further analysis.
In the case where an emission line is found with similar flux in the 300I as well as in the 300V spectrum, the average value is given.
\begin{table}[h]
\begin{center}
\small
\begin{tabular}{ll}       
\hline
Line ID  & Line flux \\
& $\times 10^{-17}$ erg\,s$^{-1}$\,cm$^{-2}$\\
\hline
\ion{[O}{II]} $\lambda$3727.4 & 2.31 $\pm$ 0.14 \\ 
\ion{[Ne}{III]} $\lambda$3868.8 & 0.47 $\pm$ 0.17 \\ 
\ion{H}{$\beta$}\,$\lambda$4861.3 & 0.44 $\pm$ 0.20\\
\ion{[O}{III]}\,$\lambda$4958.9 & 0.94 $\pm$ 0.24\\
\hline
\end{tabular}
\normalsize
\caption{Spectroscopy results for the 300V and 300I grism. We list conservative 1$\sigma$ uncertainties. For lines that were detected in both grism spectra, a weighted average of the 
fluxes was used. Spectra are corrected for Galactic extinction.  
\label{linefluxes}}
\end{center}
\end{table}

The emission lines found in our host galaxy spectra allow us to securely determine the redshift at $z = 0.858 \pm 0.001$.
The widths of the emission lines appear not to be significantly broadened, but are consistent with the instrumental resolution, as is expected from
spectroscopy of other host galaxies and the spectral resolution that was used .    

The ratios of the measured fluxes [OIII]/H$\beta$ and [OII]/H$\beta$ compared to those from samples of AGN and normal galaxies (Kennicutt 1992)
show that the host galaxy spectrum is likely not dominated by (non-thermal) AGN emission. 
We do not significantly detect the H$\gamma$ line, with an upper limit lower than the expected flux based on 
the flux of the \ion{H}{$\beta$} line and case B recombination, so we are not able to derive Balmer decrement values, 
and apply no host galaxy reddening correction to the spectrum.
\subsection{Star formation rate}
To estimate the star formation rate we use the expression from Kennicutt (\cite{Kennicutt}),
\[
{\rm SFR}_{\rm [OII]} = 1.4 \times 10^{-41} {\rm L}_{\rm [OII]} \,\,{\rm M}_\odot\, {\rm yr}^{-1},
\]
where L$_{\rm [OII]}$ = 4$\pi$d$_{\rm l}^2$f$_{\rm [OII], obs}$, with d$_{\rm l}$ the luminosity distance.
Using cosmological parameters H$_{0}$ = 70 km s$^{-1}$/Mpc, $\Omega_{\rm M}$ =
0.3 and $\Omega_\Lambda$ = 0.7, 
we find  SFR$_{\rm [OII]}$ = 1.15 $\pm$ 0.08 M$_\odot$\,yr$^{-1}$. The real uncertainty in the SFR is much larger: 
the SFR$_{\rm [OII]}$ has been calibrated on the H$\alpha$ SFR through empirical methods, providing an 
additional uncertainty through the scatter in this relation; and the effects of dust obscuration is uncertain.
This estimate for the non-extincted star formation is very typical of long GRB host galaxies (Christensen et al.~2004).

The radio continuum flux of a normal (ie.~non-AGN hosting) galaxy is  
formed by synchrotron emission of accelerated electrons in supernova remnants and by free-free emission from
HII regions, see Condon (1992). It is expected that the radio continuum flux is a particularly good tracer of the recent
SFR, due to the short expected lifetime of the supernova remnants, which is $\lesssim 10^{8}$ yr. 
We use the methods by Vreeswijk et al.~(\cite{Vreeswijk01}) and Berger et al.~(\cite{BergerSFR}) to calculate an upper limit on the unobscured star formation rate.
The three WSRT observations (see Table \ref{obstable}) combined give a 3$\sigma$ upper limit to the radio continuum flux of the 
host galaxy of 63 $\mu$Jy (see Section \ref{secradio}), and from this limit 
we find a 3$\sigma$ upper limit on the
unobscured star formation of 137 M$_\odot$\,yr$^{-1}$ (using the prescription in Berger et al.~2003). 
This is a limit deeper than that obtained for most GRB host galaxies at similar redshift (Berger et al.~2003).
We note that the
conversion of radio luminosity to star-formation rate fails in cases where there is an appreciable contribution to the radio flux from 
an AGN component, though no association of a GRB to an AGN harbouring host galaxy has been seen so far.

\subsection{Metallicity}
As the metallicity of the progenitor object plays a crucial role in our 
understanding of the evolutionary process leading to GRBs (e.g. Yoon \& Langer 2005), it is vital to secure metallicity measurements for as many GRB host galaxies as possible.
In cases where this is not possible due to redshift or signal to noise constraints, we may be able to at least ascertain the possibility that the host has 
significantly subsolar metallicity. The difference in progenitor properties between short and long bursts may also be reflected in the distribution of parameters
of the host galaxies of GRBs, such as their metallicity (but see Levan et al.~2007). 

Metallicities of GRB host galaxies are in general hard to determine: the average redshift of GRBs is high, and the hosts tend to be faint.
For the high redshift sample a line of sight value for the metallicity towards (long) GRBs can be obtained by observing the hydrogen column 
through the Lyman line(s) and the columns of lines that are not easily depleted in dust (e.g. sulphur or zinc). This method
regularly provides very accurate values of the columns of elements (e.g. Fynbo et al.~2006a; Price et al.~2007; Prochaska et al.~2007), but as the data are obtained over a line of sight and GRBs influence their environments profoundly, 
the metals and hydrogen are not necessarily co-located (e.g. Watson et al.~2007). 
High quality measurements of oxygen abundance using electron temperatures ($T_e$), determined through the [\ion{O}{III}]$\lambda$4363 line, 
can generally only be obtained for the very lowest redshift subsample of GRB hosts, and has so far only been possible for four
cases (Prochaska et al.~2004; Hammer et al.~2006; Wiersema et al.~2007). 
For GRB hosts at higher redshift, this line is often too faint to detect in a reasonable exposure time (except in cases of very high ionization parameter),
 and secondary indicators of metallicity are required.
Extensive studies have been performed to find metallicity diagnostics that depend solely on bright emission lines and on their calibration with $T_e$-derived measurements.
Generally the requirements of a secondary metallicity indicator are that the lines that are used are bright; that the correlation (based on local galaxies)
with $T_e$-determined metallicities has a low dispersion; that the line fluxes can preferably be obtained in a single spectrum (ie lines that are close together in
wavelength) and that preferably both high and low ionisation lines are used so the ionization parameter can be found.   
One of the most frequently used metallicity diagnostics using the brightest lines is the $R_{23}$ method (Pagel et al.~1979). This method gives a 
low and a high metallicity solution for a value of $R_{23}$. This degeneracy can be broken through other lines, that are often fainter and more 
difficult to detect, and the ionization parameter needed to get a good value through $R_{23}$ can be hard to determine in the case of high redshift
 and low signal to noise data. 
In the case of the host of GRB 040924, the \ion{[O}{III]}$\lambda$5007 is located right on a skyline, which makes it impossible to 
measure its flux. Instead we use a fixed ratio of \ion{[O}{III]} $\lambda$5007 to $\lambda$4959 of 3 (Izotov et al.~2006) to estimate
the [O\,III] $\lambda$5007 at $\sim$2.8 $\times 10^{-17}$ erg\,s$^{-1}$\,cm$^{-2}$.
We find $R_{23} \sim 13.8$ and using the Kewley \& Dopita (2002) formulas we find that the host flux ratio is located near the connection of the low and high 
metallcity branches, having a metallicity 12 + log(O/H)$\sim8.5$, but is badly constrained 
due to the large uncertainty in $R_{23}$, which is dominated by the uncertainty in \ion{H}{$\beta$}. It has been noted by several authors 
that the $R_{23}$ can give results discrepant from direct ($T_e$) measurements, and several empirical 
calibrations have been offered (e.g. Nagao et al.~2006). We find 
a value 12 + log(O/H)$\sim8.1$ from the $R_{23}$ value of the host of 040924 when using the Nagao et al.~calibrations.

More accurate secondary methods (e.g. involving the \ion{[N}{II]} and \ion{[S}{II]} lines)
can not be used for this host, as the diagnostic lines are redshifted out of the spectral range. 
  \begin{figure}
  \centering 
  \includegraphics[width=8.5cm]{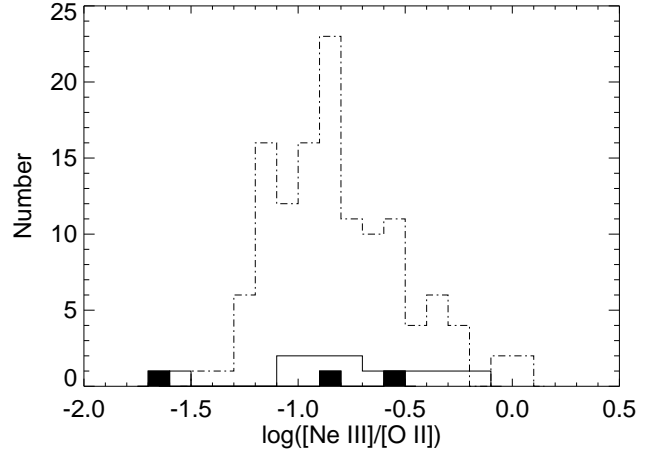}    
  \caption{Histogram showing the distributions of $\log\left(\ion{[Ne}{III]}/\ion{[O}{II]}\right)$:
   a dot-dashed line shows the SDSS spectrophotometric catalogue values 
          of sources with a determined $T_e$ metallicity from Izotov et al.~(2006), 125 sources. With a solid line the values for 
	  14 GRB host galaxies are shown. 
	  In both cases we only use sources where both lines are detected.
	  The filled squares show the values for the
	  three regions in the host galaxy of GRB\,980425 as determined by Hammer et al.~(2006), where the highest value is the bright
	  star-forming region where many WR stars are present, and the middle value is taken at the GRB location, see 
	  Hammer et al.~(2006) for details.
	  GRB hosts used are 970228 (Bloom et al.~2001), 970508 (Bloom et al.~1998), 970828 (Djorgovski et al.~2001), 980613 
	  (Djorgovski et al.~2003), 
	  990712 (Vreeswijk et al.~2001), 000418 (Bloom et al.~2003), 020405 (Price et al.~2003), 020903 (Hammer et al.~2006), 
	  030329 (Gorosabel et al.~2005), 
	  031203 (Prochaska et al.~2004), 040924 (this work), 
	  050824 (Sollerman et al.~2007), 060218 (Wiersema et al.~2007), 060505 (GRB position, Th\"one et al.~2007) and 060912A (Levan et al.~2007, priv.~comm.).}
         \label{histogram}
   \end{figure}

\subsection{\ion{[Ne}{III]} in GRB hosts}
GRB host galaxies can be found up to high redshift, and in several cases spectra of host galaxies with redshifts even higher than $z\sim1$ have been obtained. At these redshifts, the 
bright \ion{H}{$\alpha$} and [\ion{O}{III}]$\lambda$5007, 4959 are shifted out of the (useful) range of (low-resolution) optical spectra, making any secondary calibrator
that requires these lines or an accurate correction for reddening impossible to use.
It has been noted (e.g. Bloom et al. 1998; Bloom, Djorgovski \& Kulkarni 2001, Djorgovski et al.~2001) that the \ion{[Ne}{III]}$\lambda$3869 line is frequently detected in GRB 
host spectra, and its strength can be used to infer the presence of very massive stars.  
The ionization 
potentials of \ion{[Ne}{III]} and \ion{[O}{III]} are very similar (41.1 eV and 35.1 eV), so we may expect to be able to use \ion{[Ne}{III]} lines as a
substitute for \ion{[O}{III]} lines when the redshift is too high. An added advantage is the fact that the \ion{[Ne}{III]}$\lambda$3869 is close
in wavelength to the [\ion{O}{II}]$\lambda$3727 line, making uncertainties in the reddening correction less pronounced when using ratios of these
blue lines.
Nagao et al.~(2006) have gathered spectroscopic data from several samples of local galaxies with a broad range of metallicities, 
to derive unbiased samples of line fluxes and metallicities, in order to calibrate commonly used secondary metallicity indicators. Their analysis shows that
at a given metallicity the ionisation parameter has a low dispersion, and that it depends strongly on metallicity. Based on their sample of galaxies Nagao et al.~propose 
to use the ratio of the \ion{[Ne}{III]} and \ion{[O}{II]} fluxes as metallicity estimators in the absence of other, brighter high ionisation lines such 
as \ion{[O}{III]} due to high redshift. This relation is of particular importance to long GRB hosts, as their redshift distribution peaks at $z \sim 2.8$ (Jakobsson et al.~2006),
long GRB hosts tend to have high ionization parameters and because the range up to $z\sim1$ is most suitable to compare with other deep galaxy surveys (Savaglio 2006 and
references therein). 
We compile a list of GRB host galaxies with published fluxes of \ion{[O}{II]} and \ion{[Ne}{III]}, and compare the 
$\log\left(\ion{[Ne}{III]}/\ion{[O}{II]}\right)$ values of the hosts with 
those of the SDSS (DR3) sample of galaxy spectra with $T_e$ derived oxygen abundances
of Izotov et al.~(2006), i.e. a sample of emission line galaxies with detected \ion{[O}{III]}$\lambda$4363. This line becomes undetectable when the 
metallicity increases above 12 + log(O/H)$\sim8.5$ (as the H\,II regions are too cool), so this is essentially a sample of metal-poor emission line galaxies.
Figure \ref{histogram} shows the resulting distributions.
A Kolmogorov-Smirnov test shows that the GRB host sample is not significantly inconsistent (probability $P = 0.23$) with the SDSS sample of galaxies with $T_e$ metallicity determinations.
This result indicates that the sample of low redshift GRB host galaxies does not significantly differ from metal poor emission line galaxies as 
a whole in a similar 
redshift range in terms of the their integrated massive star populations, though
the current GRB host spectroscopy sample is sparse and more data is required to be certain. 
The values of the host of GRB\,980425 show that conclusions can not be drawn
about the site where the GRB occurred, but only on the integrated light of the galaxy, as has been shown in much more detail in Hammer et al.~(2006) and
Th\"{o}ne et al.~(2007). 
The value $\log\left(\ion{[Ne}{III]}/\ion{[O}{II]}\right) \sim -0.7$ for the host of GRB\,040924 is not an extreme value for a GRB host, and using the
samples of Nagao et al.~(2006) we find that the host has a metallicity of 12 + log(O/H)$\sim8.1$ from this flux ratio, matching the value from 
$R_{23}$ (Section 4.2). 

\subsection{Host SED}
The host galaxy of GRB\,040924 is clearly detected in deep HST pointings (Soderberg et al.~2006; Wainwright et al.~2007). We use
the detections in the F775 and F850LP filters, and add our R and I band detections from WHT, and upper limits in K$^\prime$ and
H band from GCNs. The R band and HST bands straddle the 4000 {\AA} break, and can therefore be used to determine some parameters of the host.
We fit galaxy templates using the HyperZ program\footnote{See {\tt http://webast.ast.obs-mip.fr/hyperz/}} developed by
Bolzonella et al.~(2000). We fit using the eight synthetic galaxy templates provided within HyperZ, using the 
spectroscopic redshift. We find that the host is best fit by a starburst template, and has $M_{\rm B} \sim-18.7$, which confirms the HST results reported by
Wainwright et al.~(2007). The age of the stellar population is young (best fit $\sim$0.12 Gyr) with little or no reddening, but these parameters are not well 
constrained as the R band magnitude is uncertain and bluer bands are not available, which makes the distinction between reddening and age hard to 
establish quantitatively. The best fit parameters are very similar to the sample of long GRB hosts analysed by Christensen et al.~(2004).
The absolute B band magnitude of the host and the estimated metallicity of 12 + log(O/H)$\sim8.1$ fits well in the local luminosity-metallicity
relation (Salzer et al.~2005), though uncertainties are large.
One characteristic of long GRB hosts is the relatively high specific starformation rate (Christensen et al.~2004). 
Assuming $M^*_B = -21.1$ and using the SFR from \ion{[O}{II]}, we find that the host of GRB\,040924 has a 
SSFR $\sim$10 M$_\odot$\,yr$^{-1}$ $(L/L_{*})^{-1}$. This value matches well with the range of SSFR $\sim 5 - 12$ found for a sample of
10 long GRB hosts analyzed by Christensen et al.~(2004).

\section{An associated supernova \label{SNsection}}
We search for a supernova associated with GRB\,040924 in the images from
WHT (see section \ref{snsearch}) by searching for a bump in the late time light curve. 
Soderberg et al.~(2006) use image subtraction with HST ACS data to derive the parameters of the supernova. They acquired three epochs,
of which the first and third are in filters F775W and F850LP, and for the second epoch only F775W was used. 
Through image subtraction they find a rebrightening of the afterglow of more than a magnitude, a clear signal of a supernova bump.
 
We use the values reported in Soderberg et al.~(2006) and Wainwright et al.~(2007) for host and supernova
measurements, and add to that our ground-based data from WHT and CAHA.
We fit the full light curve of the afterglow + host galaxy + supernova  
using the method described in Zeh, Klose \& Hartmann (2004), using SN1998bw as the template supernova. 
As we only have good measurements of the SN bump in the I band, but the afterglow is determined well only 
in the R band, we use the R band afterglow as a reference light curve. The host galaxy magnitudes in I band
determined from WHT and HST agree well with each other.
We find a luminosity factor $k=0.203\pm0.202$ and a stretch factor $s=1.371\pm0.971$. 
Essentially, the SN bump is only marginally detected through this analysis. We take $k<0.4$ (equivalent to a peak magnitude 
which is one magnitude fainter than SN 1998bw) as a conservative upper limit and thus confirm the finding 
of Soderberg et al.~(2006) that the supernova associated with this burst is  
fainter than a $K-$corrected, redshifted SN 1998bw. This can not be caused solely by extinction in the host: 
the analysis of the afterglow SED 
reveals a negligible amount of extinction along the line of sight. Even correcting for the possible small amount of 
extinction, this is still the faintest GRB-associated SN detected thus far (see Ferrero et al. 2006 and 
Sollerman et al. 2007 for the complete sample). It is most similar to the SN associated with GRB 970228 (Galama et al.~2000), 
but we caution that this SN was significantly brighter even without a correction for the extinction in the host galaxy.
As the photometric calibration uncertainty adds a large amount to the WHT data uncertainties, we perform differential photometry 
with respect to bright field stars present in all WHT epochs, and find brightening with $2.2\sigma$ in I and 1.8$\sigma$ in R, showing
that while a SN 1998bw at this redshift would have easily be detected by WHT, a GRB-SN this much fainter requires higher signal to noise data.

\section{Discussion and conclusions}
It is clear that the standard differentiation between short and long bursts using solely their duration is not a clear enough way to 
also separate out the properties of their progenitors as a whole; the distributions of long and short bursts overlap significantly (Donaghy et al.~2006). 
The same goes for the properties of their host galaxies: there
are clear associations of short GRBs with elliptical galaxies devoid of recent starformation 
(e.g. the case of GRB\,050724, Berger et al.~2005),
but there are also cases of actively starforming short burst host galaxies (e.g.~Hjorth et al.~2005; Covino et al.~2005). The lack of an optical afterglow for 
many short bursts (and therefore a sub-arcsecond position) makes the association of a galaxy with a burst rather complicated (Levan et al.~2007), making the 
diagnostic power of a galaxy association less strong by selecting only the optically luminous cases.
Possible progenitor models comprise several mechanisms (e.g. compact binary mergers, SGR giant flares), each contributing to 
the observed overall short burst population. A collection of parameters that can each on its own not decisively separate out different progenitor models,
could in combination be used for population synthesis. In a sense these parameters are not much different from the parameters that constrain
long GRB progenitors. These include (but are not limited to): the luminosity function of both GRB and afterglow, requiring the redshift distribution and
jet-break determinations; the age of the dominant stellar population, determined from host galaxy SED fitting; the quantity and age of the most massive stars,
determined through WR signatures and 
high-excitation lines; the offset of GRBs from the hosts / starforming regions; the abundance structure of the host galaxies, found through afterglow and emission line
spectroscopy; the circumburst density and its structure, found from afterglow SED modelling; the jet opening angle 
distribution, etc. 
The sample of bursts that seem to find themselves close to the classic 2 second duration division are important in these studies, as they likely contain
bursts from different progenitors and allow for easy comparison with the samples of long and short bursts.

GRB\,040924 clearly finds its origin in the collapse of a massive star. The properties of the host galaxy  fall within the range of the sample of long GRB
host galaxies: the host is blue, is actively forming stars, and has significantly sub-solar metallicity.
The afterglow, albeit faint with respect to
other pre-Swift long GRBs, behaves similarly to other long GRBs. 
The supernova could have been missed if this burst had occured at significantly higher
redshift, but although there are short bursts that share one or two of the host/afterglow properties with this burst (see
e.g.~Prochaska et al.~2006 for an overview of short burst host properties such as metallicities, luminosities etc.), the combination of 
prompt emission, afterglow and host properties makes it clear that the progenitor was a collapsar. The part of prompt emission parameter space where both long and short bursts can be 
found (i.e. 
the part of the hardness - duration diagram where the two burst distributions overlap) is interesting to explore, as we may expect to find contributions from at least two
progenitor scenarios. 

\begin{acknowledgements}    
This project is partly based on observations obtained at the ESO VLT under ESO programme 073.D-0465(B), PI R.A.M.J. Wijers.
We express our gratitude to ESO for granting our request for spectroscopy. We are very grateful to the observers that performed the observations
reported here, and we thank particularly Ian Skillen for his assistance in the planning of GRB observations.
We would like to thank Arne Henden for taking a deeper photometric calibration exposure. KW and PAC thank NWO for support under grant 639.043.302.
This study is supported by Spanish research programmes
ESP2002-04124-C03-01 and AYA2004-01515.
This paper is based on observations 
made with the William Herschel Telescope operated
on the island of La Palma by the Isaac Newton Group in the Spanish
Observatorio del Roque de los Muchachos of the Instituto de
Astrofisica de Canarias; on observations made with the Nordic Optical Telescope, operated on the island of 
La Palma, jointly by Denmark, Finland, Iceland, Norway and Sweden, in the Spanish Observatorio del Roque de los 
Muchachos of the Instituto de Astrof\'{i}sica de Canarias; and on observations collected at the German-Spanish 
Astronomical Center, Calar Alto, operated jointly 
by Max-Planck Institut f\"{u}r Astronomie and Instituto de Astrof\'{i}sica de Andaluc\'{i}a (CSIC). 
The Westerbork Synthesis Radio Telescope is operated 
by ASTRON (Netherlands Foundation for Research in Astronomy) with support from the Netherlands Foundation for Scientific 
Research (NWO). 

The authors acknowledge benefits from collaboration within the Research Training Network `Gamma-Ray Bursts: an enigma and a tool', funded
by the EU under contract HPRN-CT-2002-00294.
\end{acknowledgements}

\end{document}